# WIRELESS AND BATTERYLESS SURFACE ACOUSTIC WAVE SENSORS FOR HIGH TEMPERATURE ENVIRONMENTS

T. Aubert[1], O. Elmazria[1,2], M.B. Assouar[1]

[1]Institut Jean Lamour (IJL), UMR 7198 CNRS-Nancy University 54506 Vandoeuvre lès Nancy, France
[2]Ecole Supérieur des Sciences et Techniques d'Ingénieurs de Nancy, 54506 Vandœuvre-lès-Nancy, France
e-mail : omar.elmazria@lpmi.uhp-nancy.fr

**Abstract**

Surface acoustic wave (SAW) devices are widely used as filter, resonator or delay line in electronic systems in a wide range of applications: mobile communication, TVs, radar, stable resonator for clock generation, etc. The resonance frequency and the delay line of SAW devices are depending on the properties of materials forming the device and could be very sensitive to the physical parameters of the environment. Since SAW devices are more and more used as sensor for a large variety of area: gas, pressure, force, temperature, strain, radiation, etc. The sensors based SAW present the advantage to be passive (batteryless) and/or wireless. These interesting properties combined with a small size, a low cost radio request system and a small antennas when operating at high frequency, offer new and exiting perspectives for wireless measurement processes and IDTAG applications. When the materials constituting the devices are properly selected, it becomes possible to use those sensors without embedded electronic in hostile environments (as high temperature, nuclear site, …) where no solutions are currently used. General principle of the SAW sensor in wired and wireless configurations will be developed and a review of recent works concerning the field of high temperature applications will be presented with specific attention given to the choice of materials constituting the SAW device.

## I. INTRODUCTION

Surface acoustic wave (SAW) devices are used for several years as components for signal processing in communication systems. SAW devices are for example widely used as bandpass filter and resonator in mobile phones [1,2]. Far from being confined to this single use, SAW are or may find applications in many other areas. The SAW can be used to generate movement in microfluidics leading to mixe move, and heat very low quantities of liquid in the range of nanoliter [3]. They can also be used in chemistry, where some properties of the SAW in terms of heterogeneous catalysis could be identified. Applications in the cleaning and decontamination surface (desorption controlled ejection of drops, dust and fragments of coatings ...) are also possible. Finally, SAW offer very interesting prospects in the field of sensors. It is indeed quite possible to achieve with these devices small size and very robust sensors for deformation [4,5], temperature and gas [6,7]. Also note that in addition to being small, simple and robust, these devices have the advantage of being passive (batteryless) and remotely requestable (wireless) [8,9] and are inexpensive if fabricated on a large scale.

The use of SAW devices as passive and wireless sensors allows them to operate in extreme conditions such as those with high levels of radiation, high temperatures, or electromagnetic interference, in which no other sensors can operate. This is obviously conditioned by the fact that the materials constituting the device permit. In this paper we focus our interest on SAW sensor for high temperature environment by given a review of main works performed in this field. The study will be started by explaining the operating principle of SAW sensor and particularly the wireless SAW sensors. Note that the SAW sensors capable of operating in high-temperature environments will find applications in aerospace, power, nuclear, steel industry, automotive, chemical, and petrochemical processes plants [10,14].

## II. SAW SENSOR PRINCIPLE

The SAW devices bring into use the electro-acoustic properties of piezoelectric substrates such as quartz, lithium niobate, zinc oxide, to generate surface acoustic waves. The mechanical vibrations generated propagate at velocity depending on the crystalographic direction of material and its elastic, piezoelectric and dielectric constants [1,2].

The generation of acoustic wave is done with two electrodes, also called Interdigital transducers (IDT). The IDT consist on a structure of overlapping metal fingers that is fabricated on the piezoelectric substrate by a photolithographic process (Fig. 1). When an alternative voltage is applied to the IDT input, it follows alternating cuts and horizontal and vertical expansion of underlying material, which can generate surface acoustic waves. Different propagation modes can be generated, in this case and the commonly used is the Rayleigh wave that is similar to waves that spread on the surface of the material.

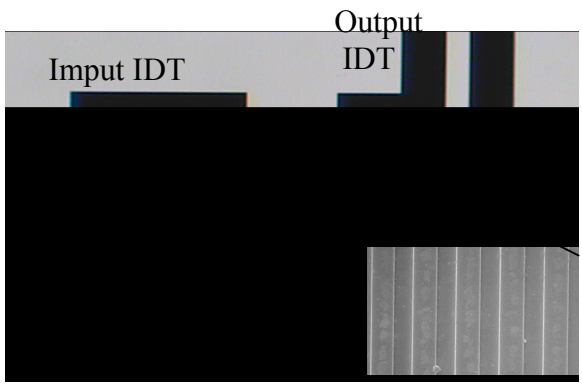

Figure 1: Optcal Photograph of SAW device (Aluuminum IDT on Quartz substrate. In insert, a scanning electron micrograph shows a zoom of the IDTs.

The acoustic wave devices are sensitive to any disturbance that may affect the velocity, distance travel or even the mode of wave propagation. A disturbance resulting in a variation of the electrical response of the device (frequency, amplitude ...). SAW systems are no exception to this rule and are sensitive to three major types of disturbances: the change in temperature, deformation and in gaseous, liquid or solid species deposited on acoustic wave travel surface. The change in temperature and deformation induces an effect on both a variation of speed (alteration coefficients elastic and piezoelectric) and a change of length to cross.

The gas species deposition (adsorption/ absorption) varies the speed of waves. This variation may result from a change of inertia mechanical surface (increase in a move masse), a modification of elastic coefficients (resulting from the spread of chemical species adsorbed in the propagation media) or a disruption of the electric field surface. In the case of the liquid or solid species their deposition affect and modify mode of wave propagation. Adding a layer of liquid or viscoelastic film on the surface may cause the emergence of guided modes at frequencies. The velocity of these modes depends on the physical parameters of the guide layer and it is therefore possible to measure them by monitoring the evolution of the frequency response of the devices. These parameters are the viscosity, density, stiffness, thickness, etc.

All SAW sensors on the market operate one or a combination of these three sensitivities.

## III. WIRLESS SENSOR PRINCIPLE

Two principles could be considered to use SAW devices as wireless sensor, delay lines or resonators. Both principles are described below:

III-1 Delay line

The operating principle of such a sensor system is sketched in figure 2. The reader unit (local radar transceiver) sends out a radio frequency (RF) electromagnetic read-out signal. This read-out signal is picked up by the antenna of the passive SAW transponder and conducted to an IDT. The IDT converts the received signal into a SAW signal by the converse piezoelectric effect. The SAW propagates towards several reflectors distributed in a characteristic pattern. A small part of the wave is reflected at each reflector. The micro acoustic wave packets now returning to the IDT are re-converted into electrical signals by the IDT and retransmitted to the radar [10].

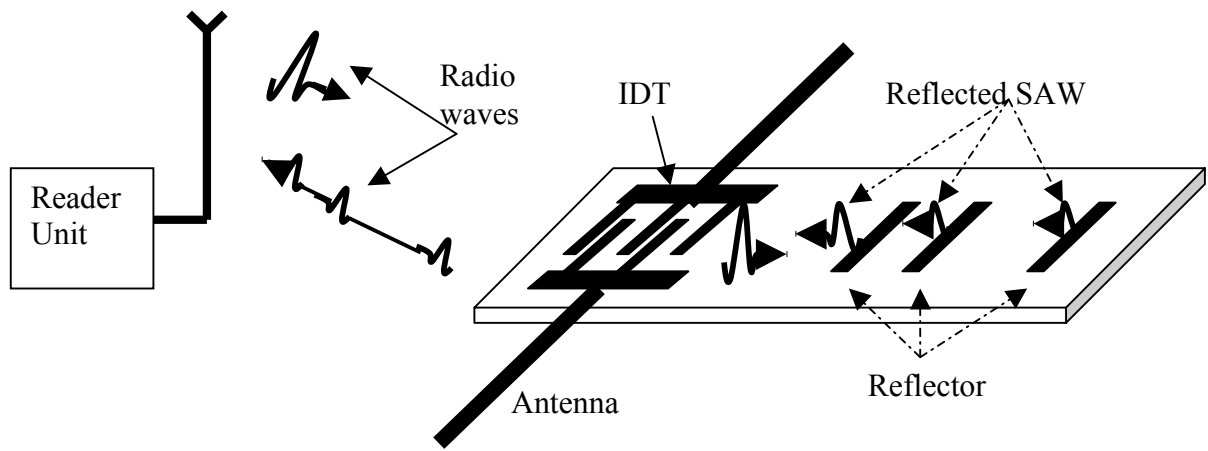

Figure 2: SAW wirless sensor in delay line configuration. Same principle is used when SAW device is considered as RFID Tag.

The mesurand (temperature, pressure, strain, …) affects the velocity of the micro acoustic wave and thereby also the time distances of the RF transponder response. The evaluation of this response signal in the radar unit thus allows the determination of the environmental temperature or pressure of the passive SAW transponder.

III-2 Resonator

The technological advances in the field of SAW have allowed the achievement resonator with a high quality factor (Q) which allows design sensor based SAW resonator (SAWR) with high sensitivity, accuracy, long-term stability and the possibility of storing electromagnetic energy,
A SAW resonator consists of the piezoelectric substrate, an interdigital transducer (IDT), and two reflectors in the direction of the propagating wave (Fig. 3).
The IDT is connected to an antenna. It receives energy for the excitation of the SAW by an electromagnetic wave coming from the interrogation unit. The IDT converts electrical energy to mechanical energy of the surface acoustic wave. The two reflector gratings form a resonating cavity in which a standing wave is generated in the case of resonance.
A portion of the stimulating electromagnetic energy is stored in this standing wave. After the stimulating signal is switched off, energy still is present in the form of the SAW. The IDT converts a portion of the mechanical energy back to electrical energy now because the process of energy conversion is partially reversible. The electrical energy is transmitted as an electromagnetic wave back to the interrogation unit and can be analyzed. Typical response obtained at the input of interrogation unit is shown in figure 4.

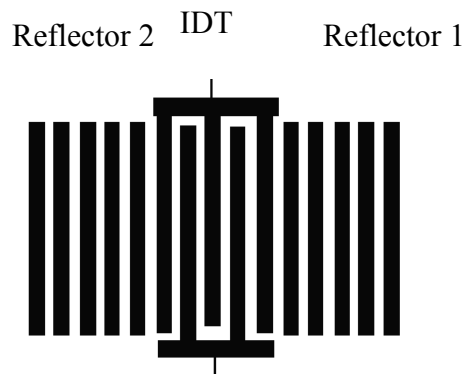

Figure 3: One port SAW resonator architecture

Duration of oscillations is of course depending on quality factor of the resonator. Resonance frequency of the sensor, directly linked to the mesurand, could be extracted from this signal using Fourrier Transform Analysis for example [11-13].

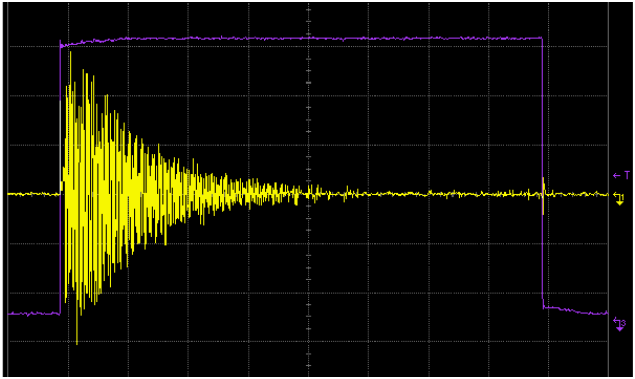

Figure 4: Typical response signal obtained at the input of interrogation unit.

## VI. SAW SENSOR FOR HIGH TEMPERATURE APPLICATIONS

As mentioned above, the use of SAW devices as passive and wireless sensors allows them to operate in extreme conditions including high temperature where no other electronic based system can operate. This unique capability makes SAW sensor very attractive for such applications. Several groups are working on achieving SAW devices capable of operating in high temperatures conditions up to 1000 °C studied for applications in aerospace, power, nuclear, chemical, and petrochemical processes plants. Knowing that the conventional substrates as piezoelectric quartz or lithium niobate are limited in temperature, R&D are focused on new generation of piezoelectric materials stable in these conditions, as Langasite (LGS) and gallium phosphate ($GaPO_4$) [10,14] or layered structures such as GaN/Sapphire, AlN/Sapphire or AlN/diamond [15]. The metal constituting the electrodes must also be resistant to such treatments and platinum (Pt, melting point 1763∘C) is preferred to, the commonly used metal aluminium (Al, melting point 660∘C), which cannot be used above 200∘C due to problems with softening and electromigration.

Additional layer is required for good adhesion between platinum and substrate. Zirconium [16] and tantalum are investigated as adhesion layer and are preferred to titanium commonly used on semiconductor application. Antenna should be also optimized for high temperature or harsh environment reached.

VI-1 Piezoelectric material

It is now well demonstrated that the conventional SAW substrates, such as quartz or lithium niobate ($LiNbO_3$) are not suitable for high temperature applications. For example, quartz knows a phase transition at 573°C, which lets him almost un-piezoelectric. However, Hornsteiner *et. al.* have even shown that SAW signals on quartz decrease dramatically above about 500°C [17]. They attribute this phenomenon to twins formation in the crystal.

In the case of lithium niobate (or lithium tantalate), the limiting factor is the chemical decomposition that starts at about 300°C [9]. Hauser *et. al.* have demonstrated that the kinetic of this phenomenon follows an Arrhenius law. Thereby, the lifetime of a $LiNbO_3$-based SAW device is about 10 days at 400°C, 1 day at 425°C, but only 2 hours at 450°C [18]. Above this temperature, they also noticed a strong deterioration of IDTs, that could be attributed to the pyroelectricity of $LiNbO_3$ which leads to the formation of sparks between the electrodes.

To overcome these limitations, new piezoelectric materials, such as langasite ($La_3Ga_5SiO_{14}$), gallium phosphate ($GaPO_4$) and aluminium nitride (AlN) have been proposed about ten years ago [19]. Since then, as we will discuss below, if langasite has paid much more attention and gave better results than gallium phosphate, aluminium nitride have not been yet really investigated in high temperature SAW applications.

- Langasite (LGS)

Langasite belongs to the same crystal class as quartz but does not undergo a phase transition up to its melting temperature at 1470°C [20]. Its electroacoustic properties at room temperature are enough good to consider it as a candidate to replace quartz in some applications. Indeed, its lower SAW velocity allowing the fabrication of smaller devices for a given frequency, the existence of temperature-compensated cuts with zero power flow angle and minimal diffraction, and especially its coupling coefficient $k^2$ which is about four times higher are strong arguments [21] (see table 1). Moreover, wafers of good quality are commercially available.

At last, elastic, piezoelectric, dielectric constants of LGS and their temperature variations up to the second order are available, allowing precise simulation [22,23]. It has to be noted that these sets are relevant in the range from 0°C to about 250°C [24,25]. There is currently no LGS constants set available for higher temperatures.

*Tab. 1 : comparison between electroacoustic properties of quartz and LGS*

|  | $v_{SAW}$ (m/s) | $k^2$ (%) | Prop. losses at 1 GHz (dB/cm) |
|---|---|---|---|
| [26] Quartz ST-X | 3200 | 0,12 | 10 |
| [9] LGS (0°, 138.5°, 26.6°) | 2700 | 0,44 | 15 |

Langasite have been intensively investigated for high temperature SAW applications in the last decade [9,17, 20-21, 25, 27-35]. The most promising results are :
- No deterioration of LGS substrates was mentioned in any of these papers
- SAW signals on langasite were measured up to 1085°C [17]. The authors pointed out this limit was that of the used package, not of LGS.
- LGS-based SAW device have been operated at 800°C for more than 5½ months [35], showing very good stability.
- LGS crystals have shown great resistance to thermal shock from ambient temperature to 700°C [35].
**All these results demonstrate that langasite is well adapted for high temperature applications, even with extreme variations of temperature.**
Nonetheless, it is necessary to slightly qualify this idyllic description. In fact, Fachberger *et. al.* have highlighted a dramatic increase of the LGS propagation losses with both the frequency and the temperature, which led them to define 1 GHz as a limit frequency for high temperature RF sensor applications [29]. The table below illustrates this problem.

*Table 2 : Maximum reached temperature for different operating frequencies*

| Operating frequency | Maximum reached temperature | Reference |
|---|---|---|
| 100 MHz | 1085 °C | [17] |
| 167 MHz | 850°C | [35] |
| 434 MHz | 750°C | [27] |
| 1 GHz | 500°C | [29] |

- Gallium phosphate

Gallium orthophosphate (GaPO$_4$) is also a promising material for high temperature electroacoustic applications. Though, its properties are a notch below those of LGS. As the latter, it has the same crystallographic structure as quartz, but it undergoes a phase transition at 930°C [36]. We can therefore expect some problems starting at about 850 °C from the experience conducted with quartz, which suffer the same type of crystalline phase transition. To our best knowledge, such temperatures have not been reached yet in acoustic applications. Buff *et. al.* have conducted experiments with GaPO$_4$-based SAW resonators up to 700°C [28], whereas some other studies on BAW and SAW devices have reached 600°C [14, 37, 38].
Anyway, electroacoustic properties of GaPO4 are slightly lower than those of LGS (see table 3). Note that this problem (strong propagation losses) is largely due to the fact that currently there are no good quality crystals available on the market [9].

*Table 3 : comparison between electroacoustic properties of GaPO$_4$ and LGS [9]*

|  | $v_{SAW}$ (m/s) | $K^2$ (%) | Prop. losses at 1 GHz (dB/cm) |
|---|---|---|---|
| GaPO4 (90°, 5°, 0°) | 2500 | 0,29 | 25 |
| LGS (0°, 138.5°, 26.6°) | 2700 | 0,44 | 15 |

- Aluminium nitride (AlN)

Aluminium nitride brings many qualities that make it a material of great interest for electronic and opto-electronic applications. It has for instance a large bandgap of 6,2 eV, high resistivity of $10^{13}$ Ω.cm and high thermal conductivity of 320 W.mK$^{-1}$ [39, 40].
In the SAW field, it has attracted considerable attention because of its high phase velocity, close to 5700 m.s$^{-1}$ [41], allowing the fabrication of high frequency devices up to 5 GHz [42, 43] when combined with high velocity substrate as diamond or sapphire. Unlike piezoelectric materials cited previously, it cannot be obtained by crystal growth methods, but must be deposited on a substrate. Several methods have been applied to obtain thin

films of AlN: MOCVD [44, 45], pulsed laser deposition [39], magnetron sputtering [46], etc

In order to use in high temperature conditions, sapphire could be a reasonable choice of substrate. Indeed, it is commercially available and not very expensive unlike diamond for example, can withstand very high temperature (at least 1600°C) [47], has a reasonable lattice mismatch with AlN allowing an epitaxial growth [48], and its surface acoustic wave velocity is very close to that of AlN [49], ensuring operating at high frequencies for AlN/Sapphire SAW devices.

The AlN/Sapphire layered structure has been intensively investigated for room temperature SAW applications because of the possibility to obtain zero TCF SAW devices [44, 45, 49]. Compared to LGS, AlN/Sapphire structure exhibits higher properties concerning SAW velocity, electromechanical coupling or acoustic propagation losses (see table 4).

*Tab. 4 : comparison between electroacoustic properties of AlN/Sapphire and LGS*

|  | $v_{SAW}$ (m/s) | $k^2$ (%) | Prop. losses at 1 Ghz (mdB/λ) |
|---|---|---|---|
| AlN (0001)/ Sapphire (0001) | 5700 [44] | 0,65 [49] | 0,7 [45] |
| LGS (0°, 138.5°, 26.6°) | 2700 [9] | 0,44 [9] | 4 [9] |

Aluminium nitride can be used in high temperature conditions. It melts at 3214°C but starts to decompose chemically in vacuum at 1040°C following the reaction $2AlN_{(s)} = 2Al_{(s)} + N_{2(g)}$ [50]. In an $N_2$-rich atmosphere (like air), the equilibrium will be shifted towards AlN formation, and the decomposition temperature will be, of course, higher. Thus, Pattel and Nicholson have measured a piezoelectric activity for AlN up to 1150°C [51]. **Moreover, the bilayer structure offers also an outstanding advantage for harsh conditions. Indeed, it is possible to sandwich the IDTs between the thin film and the substrate, offering them a natural protection against external aggression**.

Nonetheless, there are two potential limitations to the use of AlN at high temperatures. First, it is pyroelectric [19]. Currently, there is no report pointing out problems occurring at high temperature with AlN-based SAW devices due to this phenomenon, and for a good reason : to our best knowledge, **the maximum temperature of investigation for this kind of SAW device is only 350°C [52]** ! Recently, Kishi *et. al.* have successfully investigated AlN thin films as pressure sensor for engine combustion chamber up to 700°C without mentioning problems due to pyroelectricity [53]. The second problem is related to the oxidation of AlN by the reaction $4AlN_{(s)} + 3O_{2(g)} = 2Al_2O_{3(s)} + 2N_{2(g)}$. In air, it starts at 950°C and films of 500 nm thickness are completely oxidised at 1000°C in a time of minutes [54].

To summarize, conventional SAW substrates cannot be used for high temperature applications. Langasite have proved its efficiency in harsh conditions up to 1000°C, but suffers from large propagation losses at high frequency, which limit its use at frequencies below 1 GHz. This may be a problem for achieving wireless sensor with reasonable size of antenna. An alternative solution could come from the layered structure AlN/Sapphire that can theoretically operate at frequencies up to the ISM band at 2,45 GHz and up to 950°C in air.

VI-2 Metallic material for electrodes

The 3 main requirements concerning the material constituting the IDT are:
- High electrical conductivity;
- Elevated melting temperature, largely higher than the aimed operating temperature of the device (as we will discuss further)
- Good resistance to oxidation at high temperature, and more generally good chemical inertness even at high temperature

Thus, noble metals (gold, palladium, platinum, rhodium, ruthenium, iridium …), well known for their chemical inertness, are serious candidates. Melting temperature and electrical resistivity of those materials are given in table 5.

*Tab. 5 : Properties of noble metals*

| Metal | Melting temperature (°C) [55] | Resistivity (μΩ.cm) [17] |
|---|---|---|
| Gold | 1063 | 2,3 |
| Palladium | 1550 | 10,5 |
| Platinum | 1773 | 10,6 |

| Rhodium | 1966 | 4,7 |
|---------|------|-----|
| Ruthenium | 2334 | 7,7 |
| Iridium | 2440 | 5,3 |

Nonetheless, except platinum, all these metals have some weaknesses. Thus, gold has a melting temperature too close from the aimed range (400-1000°C). The solubility of oxygen is very high in palladium at high temperature [56]. Iridium, ruthenium and osmium form oxide phases at temperatures close to 700-800°C, and these oxides become volatile in the range 900-1000°C [56]. At last, Rhodium form also an oxide phase, but protective. Platinum is an exception, because the very thin (some nm) surface oxide existing at room temperature disappears at temperatures above 400°C [57]. So, platinum is the only noble metal for which no mass variation is detected up to 1000°C when heated in air [56]. This aspect is very important because any mass variation would interfere with temperature sensing, both mass and temperature variation resulting in frequency shifting of the device [28, 37]

These properties, but also the greater experience in deposition and processing of platinum over other noble metals, can explain that currently, all the studies on high temperature SAW devices were conducted quasi-exclusively with platinum electrodes [17, 20-21, 25, 27-28, 30-35]. To our best knowledge, the only exception can be found in [30] where palladium IDTs were investigated. But, the aim of the authors was to use the high $H_2$-absorption capacity of palladium to realize a SAW gas sensor.

- Adhesion layer

However, the chemical inertness of platinum (and other noble metals) have a major drawback : it can't form chemical bonds with nitride or oxide materials such as quartz, lithium niobate, langasite or aluminium nitride. As a result, the adhesion of platinum on these substrates is very poor [58]. A classical solution consists in placing underneath the platinum film a more reactive metal as adhesion layer [59, 60]. Bonding between platinum and this metal will be assured by interface alloying. Detailed studies on the adhesive strength of metals show that the optimal thickness of the adhesion layer is 10-20 nm [61]. Of course, for high temperature applications, this metal must be refractory. So, good candidates could be titanium, zirconium, tantalum, tungsten, cobalt, chromium ... In practice, titanium [14, 17, 27, 37, 55], zirconium [14, 21, 30-33, 35, 37] and tantalum [25, 34, 58, 59] are commonly used for intermediate-to-high temperature applications.

Maeder *et. al.* have compared the behaviour of these 3 adhesion layer materials after an annealing of 10 minutes in a low-pressure (10 Pa) $O_2$ atmosphere at 620°C [62] . **They showed the superiority of tantalum and zirconium over titanium**.

Indeed, Zr and Ta form dense oxide ($ZrO_2$ and $Ta_2O_5$) layers below the platinum during the annealing. It is due to the fact that $O_2$ diffuses more quickly than Zr and Ta in platinum. The important fact is that this oxidation is not a real problem, the adhesion of the platinum film remaining very good after annealing.

Titanium in turn raises serious problems already at this temperature. Diffusion of oxygen through platinum film leads also to the formation of $TiO_2$ under the latter, but unlike Zr and Ta, Ti diffuses through its oxide phase and through platinum, leading to the formation of oxide precipitates in grain boundaries inside the platinum film or near to the surface. The result is a loss of adhesion for the platinum film, the modification of its electrical properties and eventually its destruction.

Several studies have confirmed the unsuitability of titanium as adhesion layer above 600°C, as well as the good performance of tantalum and zirconium up to 700°C [14, 16, 21, 27, 30, 32, 37, 58, 59]. Thereby, AES measurements have highlighted that Zr have not diffused at all through Pt after an 8h 700°C annealing [16]. Tiggelaar *et.al.* have obtained the same result with tantalum after a 20 min annealing at the same temperature [58].

- Limitations of platinum thin film

Nonetheless, a progressive deterioration of thin films of platinum is systematically observed when they are investigated at temperatures above 700°C, even when Zr or Ta adhesion layers are used [32, 35, 58, 59]. This phenomenon starts by the apparition of crystallites at the surface of the film as shown by SEM micrograph (Fig. 5).

If the experience goes on, voids appear and grow, resulting in the formation of separate metal islands (Fig. 6) and leading to the loss of the electrical continuity of the film [58, 59] and the failure of the

SAW device. Understanding the origin of these crystallites is essential to consider alternative solutions. The most probable assumptions are :
(1) melting of a Pt-Ta or Pt-Zr eutectic
(2) alloying between Pt and Ta (or Zr)
(3) oxidation of the Pt thin film
(4) recrystallisation of the Pt thin film

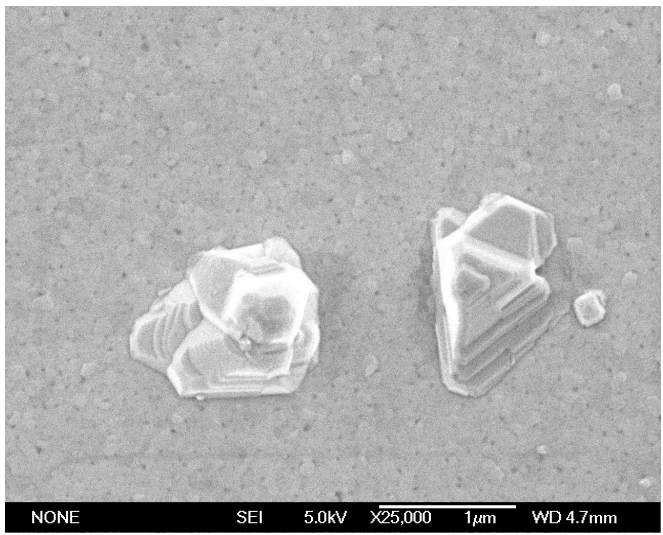

*Figure 5 : Crystallites at the surface of Pt thin film*

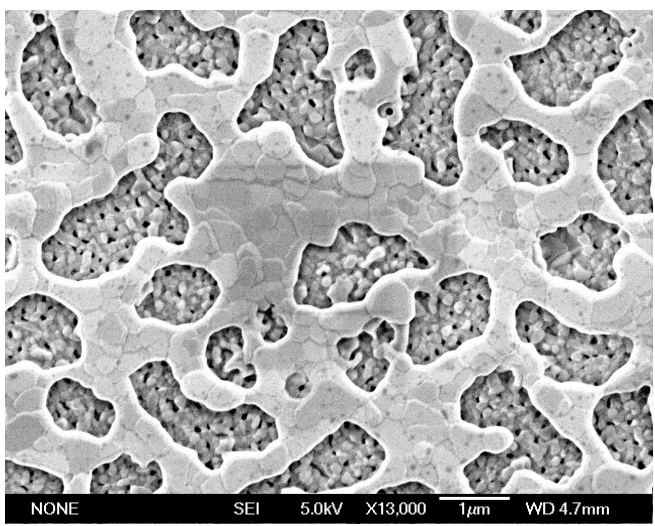

*Figure 6 : Voids at the surface of Pt thin film*

Assertions (1) and (2) can be excluded for two reasons : first, the lowest eutectic point found in the Pt-Ta diagram is at 1760°C (1190°C for the Pt-Zr case) [63]. Otherwise, Tiggelaar *et.al.* have developed a method to obtain highly adhesive Pt thin film **without adhesion layer** [58]. Now, the same phenomena of deterioration were observed with these films during annealing.

Assertion (3) can also be removed because experiments conducted in the vacuum, nitrogen or argon have leaded to the same results [32, 33, 58, 59]. Furthermore, platinum oxide $PtO_2$ cannot exist above 400°C [57].

The only remaining possibility (recrystallisation of platinum) is confirmed by XRD measurements. Indeed, the Pt peak width after annealing is about twice smaller than before. This indicates a strong increase of the Pt grain size, i.e. **an important recrystallisation** [58, 59].

- Alternative solutions

The mobility of atoms at the surface of a metal (which is essential to the recrystallisation in the case of thin films) depends on the temperature. As a rule of thumb, it is known that the latter becomes high above a "critical temperature" close to half the melting temperature of the metal (in K) [58, 59]. In the following table is presented this critical temperature for the noble metals family.

*Table. 6 : Melting and Critical temperatures of noble metals*

| Metal | Melting temperature (°C) | Critical temperature (°C) |
|---|---|---|
| Gold | 1063 | 395 |
| Palladium | 1550 | 638 |
| Platinum | 1773 | 748 |
| Rhodium | 1966 | 847 |
| Ruthenium | 2334 | 1031 |
| Iridium | 2440 | 1084 |

The correlation between the critical temperature of platinum and the experimental temperature of recrystallisation (about 700°C) is striking. Anyway, it seems likely that with gold, recrystallisation problems would occur at temperature close to 400°C. That is why gold, despite a melting temperature above 1000°C, is not used for high temperature applications. Therefore, it appears that two alternative solutions are possible. The first one would consist to use ruthenium or iridium instead platinum. In this case recrystallisation problems should not occur below

1000°C. This assumption is confirmed by [55] where an Ir thin film annealed for two hours at 1000°C in nitrogen showed excellent stability.

Nonetheless, the use of ruthenium or iridium in air at high temperature will be difficult. Indeed, these two materials are known to form volatile oxide phases at about 900°C, this phenomenon being much faster in the case of ruthenium [56]. Thus, Lisker *et.al.* have shown by XRD measurements that 100 nm-thick iridium thin film is almost completely oxidized after a 30 min annealing in oxygen at 800°C [64]. The situation is even worse with ruthenium for which quasi-complete oxidation occurs from 700°C, and high losses of the volatile $RuO_2$ phase are observed at 900°C. This effect is not observed in this work for $IrO_2$ but Krier *et.al.* have measured a rate of weight loss for iridium at 1000°C equal to 3.1 $mg/cm^2/h$, which corresponds to a recession rate of about 300 nm/h [56] ! **Therefore, it is clear that it is not possible to consider iridium (or ruthenium) IDTs without a protective layer**. **Actually, Iridium IDTs sandwiched in the AlN/Sapphire structure discussed in the piezoelectric material section could be a good solution to achieve high frequency, long-term operation wireless sensor up to 900°C.**

The second alternative solution could come from the use of Pt-Rh alloy in replacement of platinum. Indeed, rhodium, as platinum, does not undergo weight loss due to oxidation up to 1000°C [56]. Moreover, in alloys, precipitates forming in grain boundaries contribute to prevent grain boundary diffusion, and thus recrystallisation [59]. This way have been recently explored by Da Cunha *et. al.*, who have highlighted the excellent stability of a Pt/10%Rh alloy up to 950°C [31].

## V. CONCLUSION

Wireless SAW sensors are a promising solution for applications in harsh environment, such high temperature, in which no other wireless sensors can operate. However, conventional SAW substrates cannot be used for high temperature applications. Langasite have proved its efficiency in harsh conditions up to 1000°C and shows the optimal solution for wired solution. However, LGS suffers from large propagation losses at high frequency, which limit its use at frequencies below 1 GHz. This may be a problem for achieving wireless sensor with reasonable size of antenna. An alternative solution could come from the layered structure AlN/Sapphire that can theoretically operate at frequencies up to the ISM band at 2,45 GHz and up to 950°C in air. Regarding the metallic electrodes, it has been proved that platinum thin films cannot be used above 700°C for long-time applications because of recrystallisation phenomena. Interesting alternative solutions could come from the use of iridium or platinum/rhodium alloys.